\documentclass{aastex61}

\def\p0{\phantom{0}}

\def\HII{\hbox{H\,{\sc ii}}}

\newcommand{\D}{$^\circ$}

\newcommand{\RX}{\mbox {RX\,J0852.0--4622}}
\def\p0{\phantom{0}}

\received{November 14, 2017}
\revised{August 29, 2018}
\accepted{September 6, 2018}
\submitjournal{ApJ}

\shorttitle{A Morphological Study of Vela Jr.}
\shortauthors{Maxted et al.}
\begin{document}

\title{A Morphological Study of the Supernova Remnant \RX\ (Vela\,Jr.)}

\correspondingauthor{Nigel I. Maxted}
\email{n.maxted@unsw.edu.au}

\author{Nigel I. Maxted}
\affil{The School of Physics, The University of New South Wales, Sydney, NSW, 2052, Australia}
\affil{Western Sydney University, Locked Bag 1797, Penrith South DC, NSW 1797, Australia}%

\author{M. D. Filipovi\'c}
\affil{Western Sydney University, Locked Bag 1797, Penrith South DC, NSW 1797, Australia}

\author{H. Sano}
\affil{Department of Physics, Nagoya University, Furo-cho, Chikusa-ku, Nagoya 464-8601, Japan}

\author{G. E. Allen}
\affil{MIT Kavli Institute for Astrophysics and Space Research, 77 Massachusetts Avenue, NE83-557, Cambridge, MA 02139, USA}%

\author{T. G. Pannuti}
\affil{Space Science Center, Morehead State University, 235 Martindale Drive, Morehead, KY 40351, USA}

\author{G. P. Rowell}
\affil{School of Physical Science, The University of Adelaide, North Terrace, Adelaide, SA 5005, Australia}

\author{A. Grech}
\affil{Western Sydney University, Locked Bag 1797, Penrith South DC, NSW 1797, Australia}

\author{Q. Roper}
\affil{Western Sydney University, Locked Bag 1797, Penrith South DC, NSW 1797, Australia}

\author{G. F. Wong}
\affil{The School of Physics, The University of New South Wales, Sydney, NSW, 2052, Australia}
\affil{Western Sydney University, Locked Bag 1797, Penrith South DC, NSW 1797, Australia}

\author{T. J. Galvin}
\affil{Western Sydney University, Locked Bag 1797, Penrith South DC, NSW 1797, Australia}

\author{Y. Fukui}
\affil{Department of Physics, Nagoya University, Furo-cho, Chikusa-ku, Nagoya 464-8601, Japan}

\author{J. D. Collier}
\affil{Western Sydney University, Locked Bag 1797, Penrith South DC, NSW 1797, Australia}
 
\author{E. J. Crawford}
\affil{Western Sydney University, Locked Bag 1797, Penrith South DC, NSW 1797, Australia}
\author{K. Grieve}
\affil{Western Sydney University, Locked Bag 1797, Penrith South DC, NSW 1797, Australia}

\author{A. D. Horta}
\affil{Western Sydney University, Locked Bag 1797, Penrith South DC, NSW 1797, Australia}

\author{P. Manojlovi\'c}
\affil{Western Sydney University, Locked Bag 1797, Penrith South DC, NSW 1797, Australia}

\author{A. O'Brien}
\affil{Western Sydney University, Locked Bag 1797, Penrith South DC, NSW 1797, Australia}



\begin{abstract}

We conduct a multi-wavelength morphological study of the Galactic supernova remnant \RX\ (also known as Vela\,Jr., Vela\,Z and G266.2$-$1.2). \RX\ is coincident with the edge of the larger Vela supernova remnant causing confusion in the attribution of some filamentary structures to either \RX\ or its larger sibling. We find that the \RX\ radio continuum emission can be characterised by a 2-dimensional shell with a radius of 0.90$\pm$0.01$^{\circ}$ (or 11.8$\pm$0.6\,pc at an assumed distance of 750\,pc) centred at (l,b)=(133.08$^{\circ}$ $\pm{0.01}^{\circ}$,$-$46.34$^{\circ}\pm$ $0.01^{\circ}$) (or RA=8h\,52m\,19.2\,s, Dec=$-$46$^{\circ}$20$^{\prime}$24.0$^{\prime\prime}$, J2000), consistent with X-ray and gamma-ray emission. %Towards the eastern edge of \RX , filamentary radio continuum structures correspond to 
Although [OIII] emission features are generally associated with the Vela SNR, one particular [OIII] emission feature, which we denote as ``the Vela Claw'', morphologically matches a molecular clump that is thought to have been stripped by the stellar progenitor of the \RX\ SNR. We argue that the Vela Claw feature is possibly associated with \RX . 
Towards the north-western edge of \RX , we find a flattening of the radio spectral index towards another molecular clump also thought to be associated with \RX . 
It is currently unclear whether this feature and the Vela\,Claw result from interactions between the \RX\ shock and the ISM.

\end{abstract}




\section{Introduction} \label{sec:intro}
The Galactic supernova remnant (SNR) \RX\ was first discovered in 1998 \citep{Aschenbach:1998}, and studies of a tenuous detection of the radioactive decay line of $^{44}$Ti suggested \RX\ to be young ($\sim$680\,yr) and nearby ($\sim$200\,pc) \citep{Iyudin:1998}. The $^{44}$Ti detection was and remains controversial \citep{Renaud:2006}. More recent studies estimate ages of $\sim$1 to 3\,kyr and 2.4 to 5.1\,kyr \citep[][respectively]{Katsuda:2009,Allen:2015}, with distances of $\sim$700\,pc ($\pm$200\,pc). A new relevance for this object came to light when HESS observations at TeV energies \citep[HESS\,J10852$-$463,][]{Aharonian:2005,Aharonian:2007,Abdalla:2016} revealed that \RX\ is in fact a member of a class of gamma-ray SNRs that have a shell-type morphology resolved at gamma-ray energies. This reinforced the SNR's status as a key object for the study of $>$TeV cosmic-ray (CR) acceleration, which was first suggested by the presence of TeV electron acceleration \citep[][]{Slane:2001}.

A key candidate production mechanism for the gamma-ray shell of \RX\ is neutral pion production via CR interactions with gas and subsequent pion-decay \citep{Aharonian:2005,Aharonian:2007,Abdalla:2016}. Indeed, \textit{Fermi-LAT} GeV gamma-ray observations \citep{Tanaka:2011} reveal a spectrum that is compatible with cosmic-ray acceleration, but current spectral studies cannot distinguish this scenario from one where the gamma-rays are generated by a high energy electron population \citep{Abdalla:2016}, even if the spectral model is fitted to a broad wavelength range \citep[e.g. radio continuum data from][]{Stupar:2005}. 

Since neutral pion production requires the interaction of CRs with gas, attempts to identify associated gas clouds may hold the key to understanding the nature of the \RX . Recent work by \citet{Fukui:2017} has successfully identified a void in atomic gas that has a near-perfect spatial match with \RX , likely implying a core-collapse progenitor wind-blown bubble (see Section\,\ref{sec:gas}). 

\citet{Slane:2001} had previously noted that \RX\ was most likely a core collapse event, although no compact object has been conclusively associated for \RX . A coincident gamma-ray emitting Pulsar Wind Nebula (PWN), PSR\,J0855$-$4644, was identified in the south-east \citep{Acero:2013}, but it is believed to be unrelated despite having a compatible distance. A central X-ray source CXOU\,J085201.4$-$461753 at $\sim$1\,kpc was investigated as an association \citep[e.g.][]{Kargaltsev:2002}. \citet{Reynoso:2006} concluded that the object was likely unassociated with \RX, instead favouring a planetary nebula counterpart for the compact object.

Molecular gas clumps pervade the SNR boundary \citep{Fukui:2017} in a scenario where the progenitor star is argued to be associated with evaporating gaseous globules, and mirroring the molecular clumps of sister SNR, RX\,J1713.7$-$3946, which are well-studied in literature \citep[e.g.][]{Fukui:2003,Sano:2010,Maxted:2012,Maxted:2013,Fukui:2012}. \citet{Fukui:2017} find that the gamma-ray distribution traces the gas distribution, which is strong evidence for a significant hadronic gamma-ray component in the \RX\ gamma-ray emission.\footnote{An alternative view is suggested by \citet{Sushch:2018} - leptonic gamma-ray emission would also exhibit a gas-gamma-ray correlation if electron-injection becomes more efficient in parts of the SNR shock moving through high gas densities. 
Regardless, the kinematic distance solution would likely remain valid independent of gamma-ray mechanism.} 
With the newly-proposed \RX\ gas association, the time is right to test this scenario using new and archival multi-wavelength data-sets.

\subsection{The Vela SNR}\label{sec:introVela} 
The Vela\,SNR is larger and overlaps \RX , leading to the alternate names for \RX\ - Vela\,Jr. and Vela\,Z (in addition to G266.2$-$1.2). The PWN PSR\,J0855$-$4644 emits at TeV energies, but the Vela SNR itself has not been detected in HESS gamma-ray images. It follows that the Vela SNR is not an object considered for the study of $>$TeV CR acceleration. Nevertheless, the SNR may create some radio continuum structures within the \RX\ shell that may cause confusion.

The Vela SNR is foreground to Vela\,Jr at a distance of $\sim$250-350\,pc \citep[e.g.][]{Cha:1999,Dubner:1998,Caraveo:2001}. The best distance estimate probably comes from parallax measurements of the associated Central Compact Object (CCO), the Vela\,pulsar, at 287$^{+19}_{-17}$\,pc \citep{Dodson:2003pulsar}. The corresponding spin-down age \citep{Reichley:1970} of 11.4\,kyr is in agreement with the age of 18$\pm$9\,kyr, derived from the angular separation between the Vela\,pulsar and the Vela SNR centroid (as indicated by `explosion fragments') given its measured proper motion \citep{Aschenbach:1995}.

% Emission mechanism and nature, gamma velorum bubble
\citet{Sushch:2011} proposed that the evolution of Vela SNR took place inside the wind-blown bubble of the Wolf-Rayet star, $\gamma ^2$\,Velorum. The model put the Vela SNR at the east-north-eastern side of the bubble, with interactions taking place on the foreground side of the bubble. The resulting density difference between the shock component expanding into the bubble and the shock component expanding into the stellar bubble boundary might account for the asymmetry of the Vela SNR shell \citep{Sushch:2011}. \citet{Kim:2012} found this model to be consistent with the discovery of FUV filaments along regions proposed to have a higher density. The authors also noted the possible first detection of Vela\,Jr at FUV wavelengths through the examination of atomic line ratios (primarily OIII]/OIV] and OIII]/CIV]) that are more indicative of non-radiative shocks than other parts of the Vela SNR shell. The implication is that Vela\,Jr, which has non-radiative shocks, may be responsible for a component of oxygen ion emission seen by \citet[][]{Miller:1973} or \citet[][]{Nichols:2004}. This is despite low-ionisation oxygen emission generally being associated with cooling in post-shocked gas associated with older radiative shocks. 

%The Vela SNR shock has a relatively slow radiative forward shock. 
While Vela\,Jr generally exhibits shock speeds of $\sim$3000\,kms$^{-1}$ \citep[][]{Katsuda:2009}, the measured velocity of the foreground Vela SNR shock front ranges between $\sim$100 and 280\,kms$^{-1}$ \citep{Cox:1972,Raymond:1991,Jenkins:1995,Bocchino:1999,Cha:2000,Bocchino:2000,Pakhomov:2012}, while so-called `explosion fragments' exhibit velocities of 660-1020\,kms$^{-1}$ \citep{Aschenbach:1995,Sushch:2011}. \citet{Redman:2000} find optical [SII] emission from one such fragment, RCW\,37. The large associated speed and temperature lead the authors to suggest that Vela\,Jr may be responsible for the fragment feature. This is in contrast to the majority of filamentary structures seen in optical atomic and ionic emission lines in the region. Emission lines such as [OIII] have been attributed to the cooling of low-density shock-heated regions associated with shock speeds of $\sim$100\,kms$^{-1}$ in the Vela SNR \citep[e.g.][]{Raymond:1997,Sankrit:2004}.\footnote{We note that such emission is also characteristic of the similar-speed shocks of Herbig-Haro objects \citep[e.g.][]{Schwartz:1978,Hartigan:1987}.}
UV emission of higher transition emission lines are attributed to faster components (150-170\,kms$^{-1}$) within the Vela SNR shock \citep[e.g.][]{Raymond:1981,Slavin:2004,Nichols:2004,Sankrit:2004,Kim:2012}.

We conduct a morphological investigation using radio-continuum data (including radio spectral index), HI, CO, H$\alpha$, [SII], [OIII], UV, X-ray and gamma-ray emission. We attempt to attribute components within the \RX\ field of view to either \RX\ (i.e. Vela\,Jr) or the overlapping Vela SNR.

%History of discovery
%historical radio towards Vela X, then Vela Z
\subsection{An Expanded History of \RX}
First imaged in X-rays by the \textit{ROSAT} all-sky survey, \RX\ (i.e. Vela\,Jr) was initially apparent at E$>$1.3\,keV \citep{Aschenbach:1998}. Before this X-ray discovery there were no reports of a radio SNR coincident with \RX\, although many Vela and Galactic Plane surveys covered the area. \citet{Duncan:1996} and \citet{Bock:1998} surveyed the SNR region with both the Parkes radio telescope and the Molonglo Observatory Synthesis Telescope (MOST) at 2420\,MHz and 843\,MHz, respectively. Before this time, the Parkes-MIT-NRAO (PMN) survey \citep{Griffith:1993} covered the area containing \RX\ at 4850\,MHz, and a candidate non-thermal source coinciding with the north-eastern limb of \RX\ was identified at 408\,MHz as long ago as 1968 \citep{Milne:1968}.

\citet{Combi:1999} presented a reanalysis of 2420\,MHz data and a low angular resolution (30$'$) detection of Vela\,Jr at 1420\,MHz in light of its discovery overlapping the Vela SNR at keV wavelengths. The authors concluded that the overall spectral index is $\sim -0.3$ which is flatter than expected for young SNRs (typically $\alpha \sim -0.7$)\footnote{where $S_\nu \propto \nu^\alpha$}. Follow-up observations \citep{Duncan:2000} measured spectral index ($\alpha =-0.4\pm0.15$) for \RX\ that are consistently flatter than similar age SNRs in the Milky Way and Magellanic Clouds \citep[MCs,][]{Bozzetto:2017}.

\citet{Filipovic:2001} showed that the structure of \RX\ is shell-like (barrel-shaped/bilateral) with good correlation between radio-continuum, EUV and \textit{ROSAT} PSPC X-ray emission. The radio-continuum emission coinciding with X-rays confirmed that synchrotron radiation is responsible for the north brightened X-ray limb \citep{Bamba:2005,Pannuti:2010}.

\citet{Stupar:2005} presented a multi-frequency radio-continuum study of \RX\ based on low-resolution mosaic observations with the Australia Telescope Compact Array (ATCA) radio interferometer at 1384 and 2496\,MHz, Parkes 4850\,MHz and MOST 843\,MHz survey data. They determined the radio spectral index for several prominent features of this SNR and found a sudden spectral turn over at 1384\,MHz, but based on a closer inspection of the data leading to this turnover feature, we argue that observational shortcomings put the existence of this feature in doubt. This is due to the lack of short spacings within the array configuration used. In response to this, as part of this investigation we present new spectral index data towards \RX .

\section{Observational data} \label{sec:obs}
Various observations have been carried out on \RX\ over a period of 15 years. According to literature, \RX\ is centred at RA(J2000) = $8^{h}52^{m}3^{s}$ and DEC(J2000) = --46\D22$'$ \citep{Aschenbach:1998} which is within the bounds of the larger, Vela SNR. In this study, we highlight GHz radio continuum, optical emission line and ultraviolet (UV) data.

\subsection{ATCA Radio Continuum}\label{sec:cont}
The Australia Telescope Compact Array (ATCA) is an array of six 23\,m dishes in Narrabri, New South Wales in Australia. The observations that were analysed in this study were obtained from the ATNF online archive -- ATOA\footnote {http://atoa.atnf.csiro.au/}. A list containing the observations that were analysed, which span the frequency range between 1384 and 2868\,MHz, are displayed in Table \ref{tab:observations_table}. 
\begin{table*}
	\caption{Summary of the ATCA project file radio-continuum observations of \RX\ used in this study.}
	\center
	\begin{tabular*}{\textwidth}{@{\extracolsep{\fill}} l l c c c c }
		\hline
		Project & Dates        & Array & $\nu$ & Bandwidth & Channels \\
		Code    &              &       & (MHz)     & (MHz) &   \\  
		\hline
		C789  & 14-15 Nov 1999 & 210   & 2496  & 128       & 33 \\
%		C1481 & 17 Jan 2006    & EW352 & 1384  & 128       & 33 \\
%		C1481 & 17 Jan 2006    & EW352 & 1728  & 128       & 33 \\
%		C1481 & 19 Mar 2006    & EW367 & 1384  & 128       & 33 \\
%		C1481 & 19 Mar 2006    & EW367 & 1728  & 128       & 33 \\
		C2449 & 26-27 Feb 2011 & EW352 & 2100  & 2048      & 2049 \\
		C2449 & 29-30 Mar 2011 & EW367 & 2100  & 2048      & 2049 \\
%		CX310 & 29-30 Dec 2014 & 6A    & 2100  & 2048      & 2049 \\
		\hline
	\end{tabular*}
	\label{tab:observations_table}
\end{table*}

ATCA data from 1999 were taken as part of the Southern Galactic Plane Survey \cite[SGPS,][]{McClure:2005} and had full coverage of the the \RX\ shell, while more recent measurements from 2011 were from a campaign targeting HI emission in the SNRs south-west \citep{Fukui:2017}. The latter observations were comprised of 43 pointings taken in mosaic mode and arranged in a hexagonal grid that covers approximately half of the SNR. The introduction of the Compact Array Broadband Backend system \citep[CABB,][]{Wilson:2011} to ATCA provided a factor 16 increase in the observing bandwidth compared to earlier observations of \RX\ \citep{Stupar:2005,Pannuti:2010}. From 2$\times$128\,MHz to 2$\times$2048\,MHz IF bands, the addition of 16 zoom windows significantly improved the RMS noise and therefore detections of features of this SNR. With the added bandwidth and functionality from the inclusion of CABB, separate spectral line observations can be made using CABBs zoom band mode. This allowed for observations of both continuum and spectral lines simultaneously from 2011 onwards. These data were taken with an increased image-size for each pointing, facilitating effective image cleaning techniques \citep[e.g. `Peeling', as described in][]{Hughes:2006,Crawford:2011}.

\begin{figure}[h]
\centering
\includegraphics[angle=0, trim=0 0 0 0, width=0.75\columnwidth]{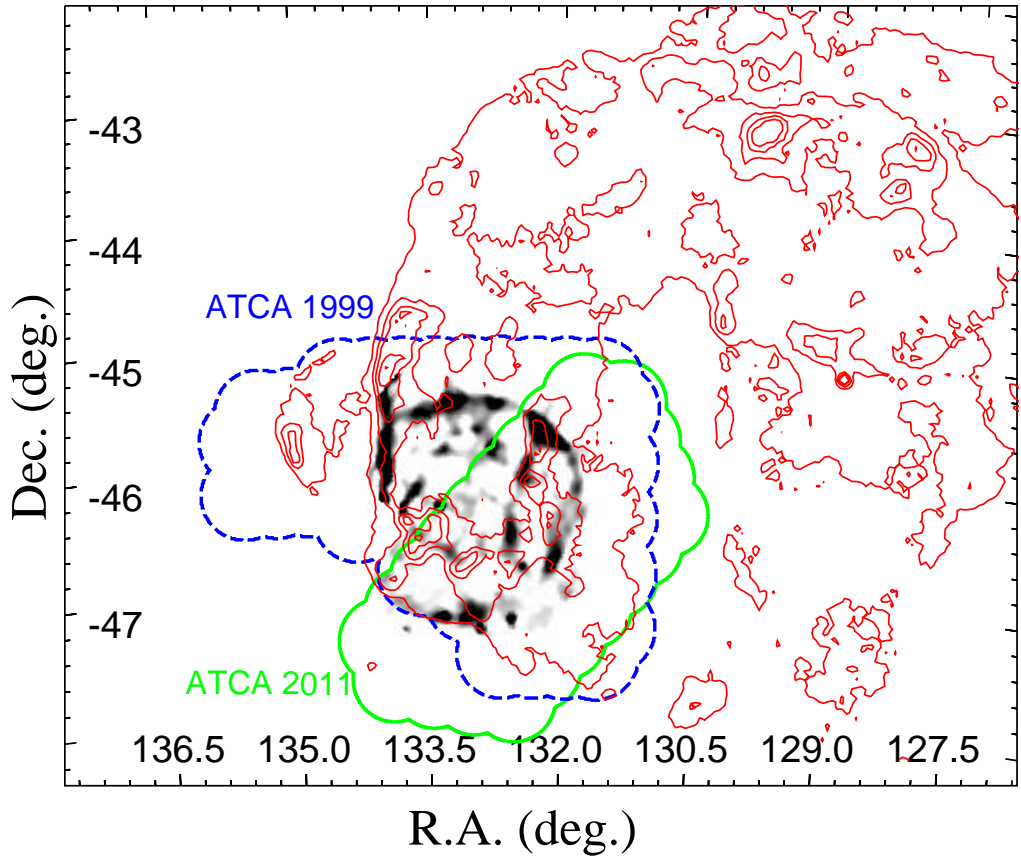}
\caption{A noise-weighted combination of ATCA radio continuum data spanning frequency bands of 1332 to 2868\,MHz. Regions of ATCA coverage from 1999 and 2011 observation campaigns are indicated by blue dashed and solid green regions, respectively. \textit{ROSAT} broad-band (0.1-2.4\,keV) X-ray count contours (5, 15, 25, 35, 45\,arcmin$^{-2}$) are overlaid, dominated by the soft thermal component of Vela SNR X-ray emission \citep{Aschenbach:1998}.
\label{fig:ATCAcoverage}}
\end{figure}

The properties of each image examined in this analysis can be found in Table \ref{tab:image_table}. Figure\,\ref{fig:ATCAcoverage} indicates the coverage of the 1999 and 2011 ATCA observation campaigns. Images were processed using the MIRIAD\footnote{http://www.atnf.csiro.au/computing/software/miriad/} software package \citep{Sault:1995}. Frequencies range between 1332\,MHz and 2868\,MHz and typical RMS noise levels are $\sim$1\,mJy/beam.
\begin{table*}
	\caption{ Summary of various image properties of \RX\ used in this study.}
	\center
	\begin{tabular*}{\textwidth}{@{\extracolsep{\fill}} h l c c c c c c }
 		\hline
Fig. & Project & Frequency & Bandwidth & Pixel Size & Beam Size & Position Angle & RMS \\
No.  & Code    & (MHz)     & (MHz)     & (arcsec)   & (arcsec)  & (degrees) & (mJy/beam) \\
  		\hline
 		\ref{fig:Rad_C789}  & C789  & 1384 & 128 & 51.4 & 247.6 $\times$ 179.4 & 60.4    &  1.5\p0  \\
 		\ref{fig:Rad_C789}  & C789  & 2496 & 128 & 28.9 & 118.8 $\times$ 89.0  & 55.4    &  1.0\p0  \\
% 		\ref{fig:Rad_C2449} & C1481 & 1384 & 128 & 29.6 & 137.1 $\times$ 109.6 & \p01.9  &  2.0\p0  \\
 		\ref{fig:Rad_C2449} & C2449 & 1332 & 512 & 27.2 & 120.4 $\times$ 99.0  & \p07.3  &  1.0\p0    \\
 		\ref{fig:Rad_C2449} & C2449 & 2100 &2048 & 14.1 & 80.8  $\times$ 69.0  & \p07.7  &  0.4\p0  \\
 		\ref{fig:Rad_C2449} & C2449 & 1844 & 512 & 20.3 & 91.6  $\times$ 77.4  & \p01.9  &  1.0\p0  \\
 		\ref{fig:Rad_C2449} & C2449 & 2356 & 512 & 16.3 & 71.9  $\times$ 61.5  & 10.9    &  1.0\p0  \\
 		\ref{fig:Rad_C2449} & C2449 & 2868 & 512 & 14.1 & 59.5  $\times$ 50.0  & \p08.7  &  1.0\p0  \\
% 		\ref{fig:Rad_CX310} & CX310 & 2100 &2048 & 1.75 & 14.6  $\times$ 7.1   & \p06.5  &  0.03   \\
  		\hline
	\end{tabular*}
	\label{tab:image_table}
\end{table*}

\begin{figure}[h]
\centering
\includegraphics[angle=0, trim=0 0 0 0, width=0.45\columnwidth]{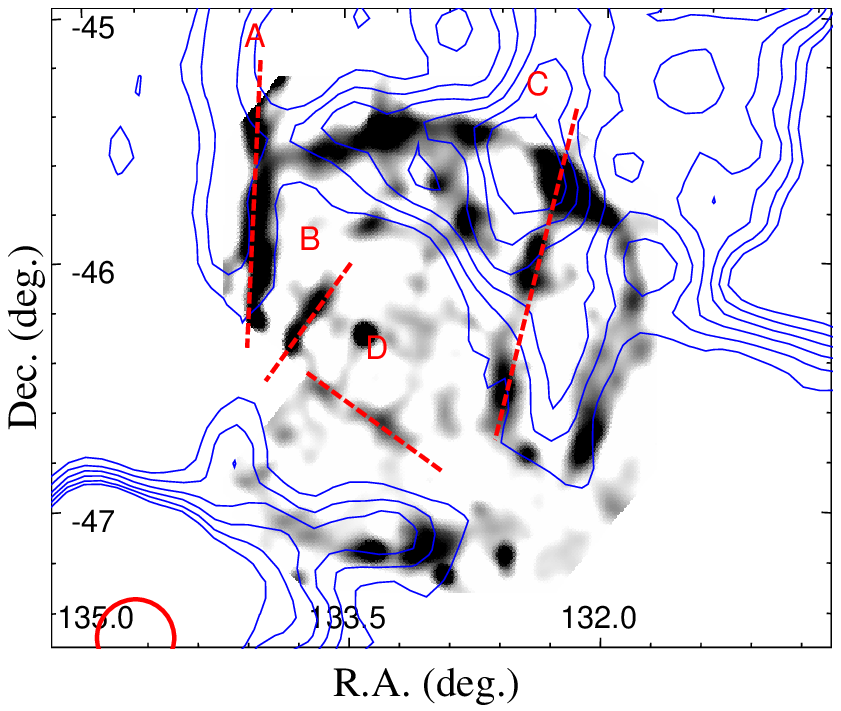}
\includegraphics[angle=0, trim=0 0 0 0, width=0.45\columnwidth]{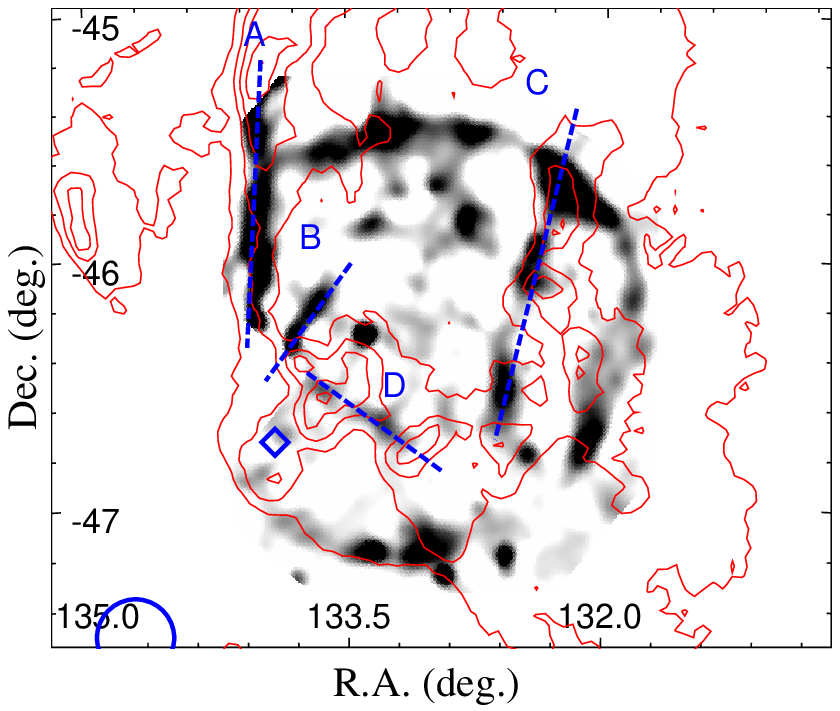}\\
\includegraphics[angle=0, trim=0 0 0 0, width=0.45\columnwidth]{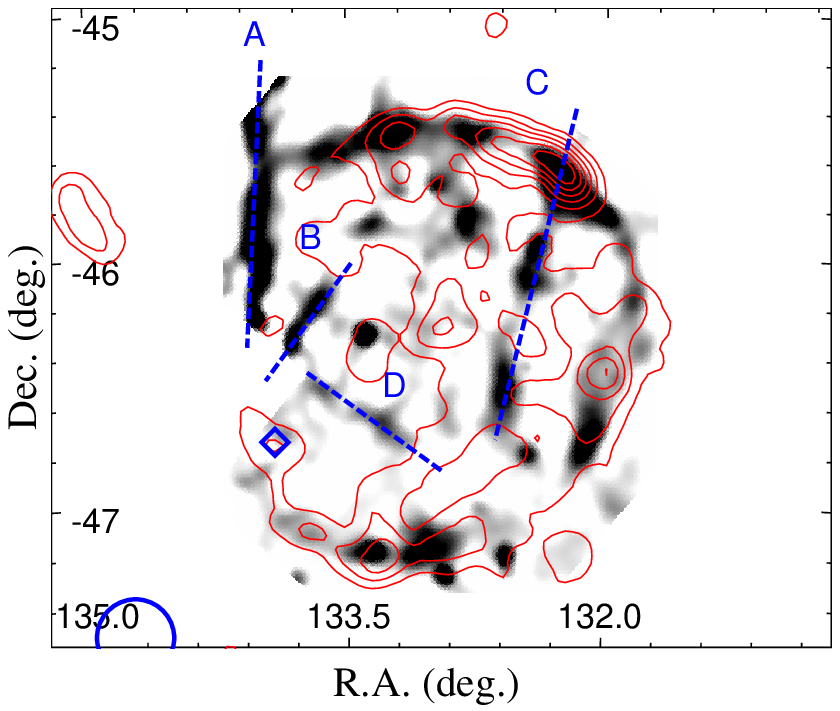}
\includegraphics[angle=0, trim=0 0 0 0, width=0.45\columnwidth]{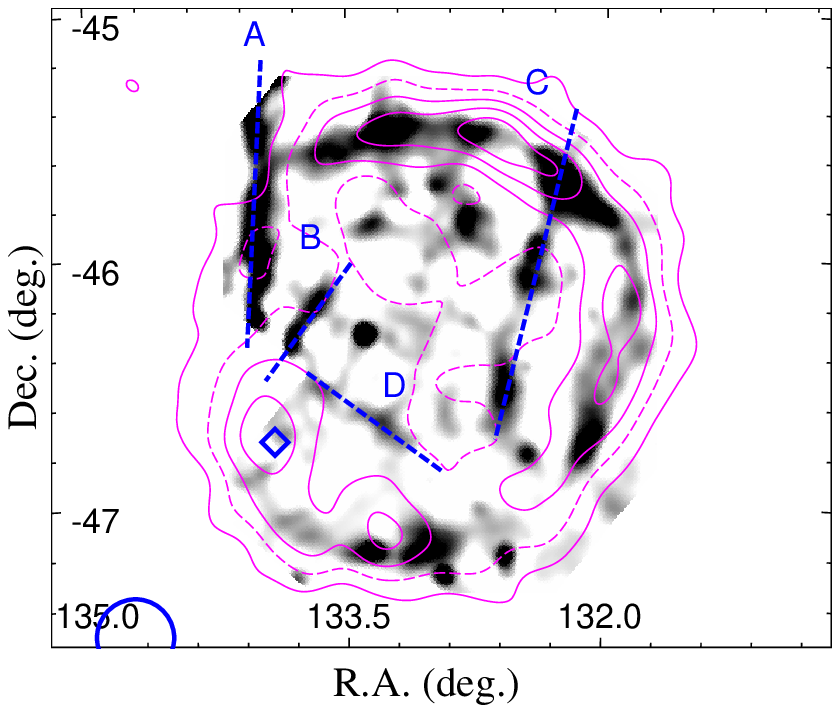}
		\caption{A noise-weighted combination of ATCA radio continuum data spanning frequency bands of 1332 to 2868\,MHz, and years 1999 to 2011 (project codes C789 and C2449). A circle indicates the position of the strong radio source, RCW\,38, and diamond indicates the position of PSR\,J0855$-$4644. Dashed lines, labelled A-D, indicate filamentary structures discussed in Section\,\ref{sec:filaments}.
In the top left image, 13\,cm Parkes radio continuum contours (0.4, 0.6, 0.8 1.2, 1.4\,Jy/beam) are overlaid \citep{Duncan:2000}. 
In the top right image, \textit{ROSAT} broad-band (0.1-2.4\,keV) X-ray count contours (5, 15, 25, 35, 45\,arcmin$^{-2}$) are overlaid \citep{Aschenbach:1998}.
In the bottom left image, \textit{ROSAT} hard ($>$1.3\,keV) X-ray count contours (0.8, 1.2, 1.6, 2.0, 2.4, 2.8) are overlaid \citep{Aschenbach:1998}.
In the bottom right image, HESS TeV gamma-ray excess count (65, 80, 95) contours are overlaid. \label{fig:Rad_combo}}
\end{figure}

Figure\,\ref{fig:Rad_combo} shows a final a noise-weighted combination of the 1999 and 2011 ATCA radio continuum data-sets (project codes C789 and C2449) spanning frequency bands between 1332 to 2868\,MHz. The image largely replicates features observed in images by \citet{Stupar:2005} and was produced to encapsulate the morphological features seen in individual images. The very bright, radio-loud \HII\ region outside of the Vela\,Jr field, RCW\,38, produced large side lobes, so image boundaries were cropped prior to merging and data-sets were normalised to ensure that morphological features were continuous. The final image, which encompasses $\sim$320$^{\circ}$ of the azimuthal angle of the SNR radio shell, is displayed in Figure\,\ref{fig:Rad_combo}. This image has limitations in coverage in the South-East and in the interpretation of flux, but can clearly illustrate filaments discussed in Section\,\ref{sec:filaments}, so is ideal for morphological studies. 

\begin{figure}[h]
	\centering
	\includegraphics[width=0.8\columnwidth]{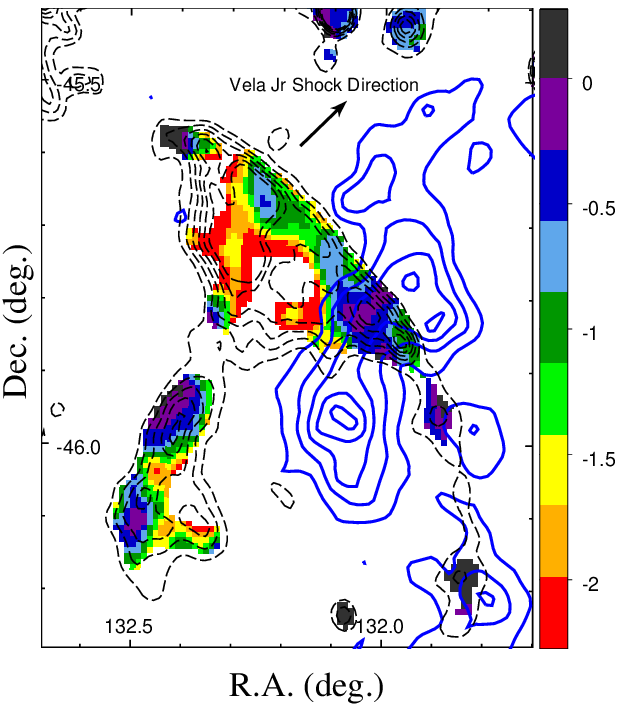}
	\caption{Spectral index derived from project C2449 (year 2011) towards the North-Western rim of \RX\ (Vela\,Jr). Black, dashed radio continuum contours from a single frequency band are overlaid. Blue 2, 4, 6, 8, 10 and 12\,K\,km\,s$^{-1}$ Nanten $^{12}$CO(1-0) integrated emission (28-33\,km\,s$^{-1}$) contours are also overlaid \citep[as seen in Figure\,1 of][]{Fukui:2017}. The estimated uncertainty in spectral index is $\sim$15-20\%.
	\label{fig:SpecIndex}}
\end{figure}

In addition to intensity images, we also present a radio spectral index map of \RX\ in Figure\,\ref{fig:SpecIndex}. 
The 1999 and 2011 data were examined independently, but only the 2011 results are displayed. The array configuration used to take the 1999 data lacked short-spacings and systematic effects were introduced into the corresponding spectral index map. 

\subsection{Optical Data} \label{sec:opt}
\begin{figure*}[h]
	\centering
	\includegraphics[width=1.0\columnwidth]{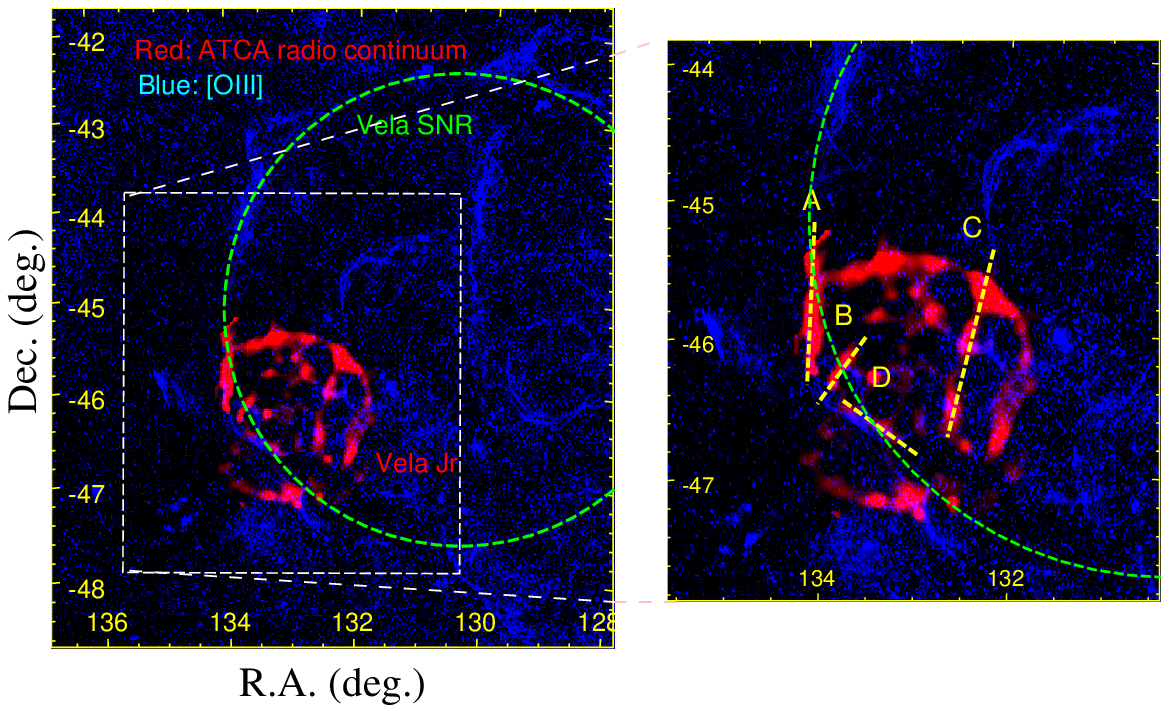}
	\caption{A 2-color image of 500\,nm [OIII] emission image \citep{Filipovic:2001} in blue and ATCA 1332-2868\,MHz radio continuum emission in red. The approximate location of the Vela SNR is indicated by a dashed-green circle. In the right-hand image, yellow dashed lines, labelled A-D, indicate filamentary structures.\label{fig:OIII}}
\end{figure*}

This study utilises measurements from the Australia National University 16\,inch Boller \& Chivens Telescope\footnote{http://rsaa.anu.edu.au/observatories/telescopes/anu-16-inch-boller-chivens-telescope} in February 1999 \citep{Filipovic:2001}. Filters targeted the doublet [OIII], H$\alpha$ and doublet [SII] transitions at 4861.3/5006.9, 6562.8 and 6718.3/6732.7\,\r{A}, respectively, over an exposure time of 600\,seconds. We display [OIII] emission in Figure\,\ref{fig:OIII}.

\subsection{UV Data}
\begin{figure}[h]
	\centering
	\includegraphics[width=0.75\columnwidth]{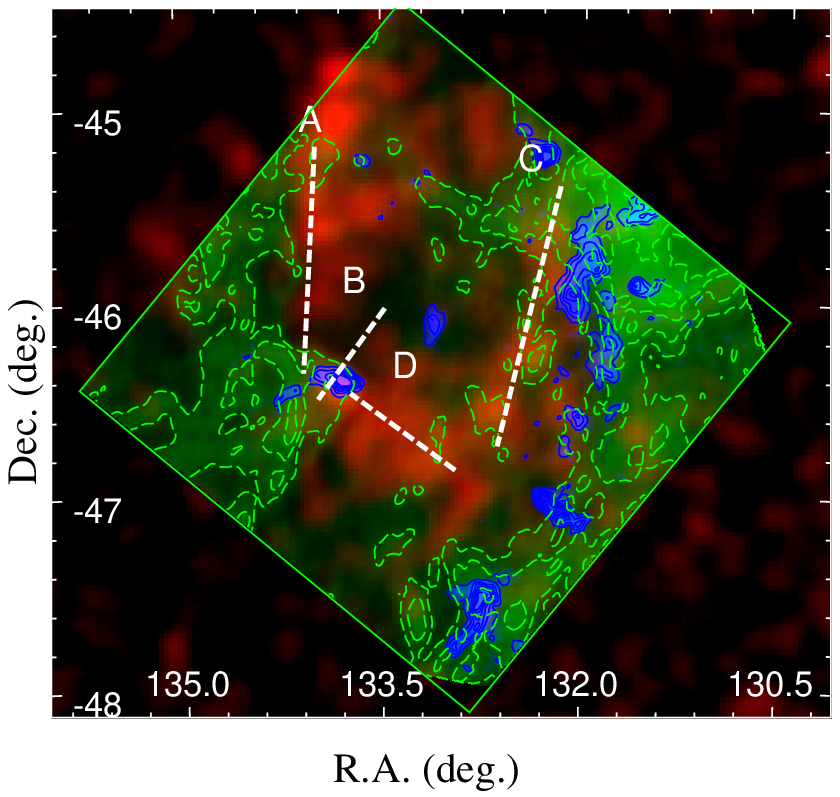}
	\caption{A 3-colour image featuring 58-174\,\r{A} Ultra Violet emission from the EUVE satellite (red), and gas tracers at the velocity of a void. ATCA and Parkes HI emission integrated between 22 and 33\,km\,s$^{-1}$ is green, with dashed green contour levels of 150, 200, 250 and 300\,K\,km\,s$^{-1}$ is displayed \citep[as seen in Figure\,1 of][]{Fukui:2017}. Nanten $^{12}$CO(1-0) emission integrated between 28 and 33\,km\,s$^{-1}$ is blue in the image, with blue 2, 4, 6, 8, 10 and 12\,K\,km\,s$^{-1}$ contours overlaid \citep[as seen in Figure\,1 of][]{Fukui:2017}.\label{fig:GasUV}}
\end{figure}

58-174\,\r{A} UV images from the Extreme Ultraviolet Explorer \citep{Welsh:1990} are displayed in Figure\,\ref{fig:GasUV}. UV data was collected using 2 Wolter-Schwarzschild Type I grazing incidence mirrors with a microchannel plate detector\footnote{https://heasarc.gsfc.nasa.gov/docs/euve/euve.html}. The average exposure across the Galaxy is generally greater than 500\,s.

The data was taken between 1991 and 1993 \citep{Edelstein:1993} and was utilised in previous studies by \citet{Filipovic:2001}. The data have a natural angular resolution of $\sim$6$\times$6$^{\prime}$, and was smoothed to $\sim$10$^{\prime}$ resolution.

\subsection{Spectral CO and HI data}
\citet{Fukui:2017} presented an analysis of the interstellar medium towards \RX\ with a focus on HI emission, and molecular clumps traced by CO(1-0). The authors identified a candidate interstellar medium association for \RX\ and their data are used in our multi-wavelength investigation.

As detailed in \citet{Fukui:2017}, 21\,cm HI data were taken with ATCA (see Section\,\ref{sec:cont}) and the 64\,m Parkes telescope. The resultant combined data-set has a beam FWHM of 245$^{\prime\prime}\times$130$^{\prime\prime}$. CO(1-0) data at 115.290\,GHz were taken with the Nanten telescope with beam FWHM of 160$^{\prime\prime}$

\section{Results and Discussion}
Like in radio continuum images by \citet{Stupar:2005}, in Figure\,\ref{fig:Rad_combo} the circular structure can be clearly discerned in the north and north-west of the \RX\ shell, in addition to the southern section. This \RX\ (Vela\,Jr) circular structure appears consistent with Parkes 13\,cm data \citep{Duncan:2000}, hard $>$1.3\,keV X-ray emission \citep{Aschenbach:1998} and TeV gamma-ray emission \citep{Abdalla:2016}, whereas, as noted in the \citeauthor{Aschenbach:1998} Vela\,Jr discovery paper, the broadband X-ray (0.1-2.4\,keV) structure does not reflect the circular structure of Vela\,Jr. 

We characterise the radio emission towards \RX\ as a circular shell in Section\,\ref{sec:shell} and highlight several filamentary radio continuum structures 
in Figure\,\ref{fig:Rad_combo} - Filament A, B, C and D. These are discussed in Section\,\ref{sec:filaments}.

\subsection{Parametrising the SNR shell}\label{sec:shell}
We perform $\chi^2$-minimisation fits of a 2-dimensional shell model\footnote{http://cxc.harvard.edu/sherpa/ahelp/shell2d.html} to each radio continuum data-set in Table\,\ref{tab:image_table} using SHERPA\footnote{http://cxc.cfa.harvard.edu/sherpa/} software. We masked rectangular regions encompassing three filamentary structures (A, B and C, see Section\,\ref{sec:filaments}) then employed a Neldermead fitting method \citep{Lagarias:1998}, assuming a constant noise level across each image. 

The centre position, SNR radius and shell width were free parameters in the fit. The best-fit parameters are displayed in Table\,\ref{tab:sherpa}. 

\begin{table*}
\centering
\caption{Results of a $\chi^2$-minimisation fit of a 2-dimensional shell function to radio continuum data. Central position (J2000 R.A.,Dec.), Radius and Width were solved for each radio continuum image. The displayed width has been deconvolved assuming a beam FWHM that is the average of the minor and major axis of the beam FWHM (see Table\,\ref{tab:image_table}).\label{tab:sherpa}}
\begin{tabular}{|l|l|l|l|l|l|l|}
\hline
year	&	Central		&	Right		&	Declination	&	Radius$^{\dagger}$	&	Width		\\				
	&	Frequency	&	Ascension	&				&			&				\\			
	&	(MHz)		&	(deg)		&	(deg)		&	(arcsec)	&	(arcsec)		\\				
\hline
1999	&	1384	&	132.98 	$_{	-0.002 	}	^{+	0.001 	}$	&	-46.26 	$_{	-0.001 	}	^{+	0.001 	}$	&	2943 	$_{	-4 	}	^{+	12 	}$	&	50 	$_{	-2 	}	^{+	13 	}$	\\
	&	2496	&	132.99 	$_{	-0.004 	}	^{+	0.003 	}$	&	-46.27 	$_{	-0.002 	}	^{+	0.003 	}$	&	2965 	$_{	-12 	}	^{+	17 	}$	&	125 	$_{	-14 	}	^{+	7 	}$	\\
\noalign{\smallskip}
2011	&	1332	&	133.19 	$_{	-0.013 	}	^{+	0.008 	}$	&	-46.29 	$_{	-0.004 	}	^{+	0.010 	}$	&	3380 	$_{	-35 	}	^{+	49 	}$	&	185 	$_{	-46 	}	^{+	12 	}$	\\
	&	1844	&	133.22 	$_{	-0.010 	}	^{+	0.013 	}$	&	-46.27 	$_{	-0.009 	}	^{+	0.007 	}$	&	3433 	$_{	-46 	}	^{+	44 	}$	&	160 	$_{	-22 	}	^{+	41 	}$	\\
	&	2100	&	133.19 	$_{	-0.009 	}	^{+	0.010 	}$	&	-46.28 	$_{	-0.008 	}	^{+	0.007 	}$	&	3377 	$_{	-48 	}	^{+	38 	}$	&	157 	$_{	-36 	}	^{+	26 	}$	\\
	&	2356	&	133.23 	$_{	-0.014 	}	^{+	0.018 	}$	&	-46.26 	$_{	-0.010 	}	^{+	0.015 	}$	&	3471 	$_{	-78 	}	^{+	85 	}$	&	152 	$_{	-46 	}	^{+	60 	}$	\\
	&	2868	&	133.23 	$_{	-0.013 	}	^{+	0.017 	}$	&	-46.26 	$_{	-0.013 	}	^{+	0.009 	}$	&	3456 	$_{	-73 	}	^{+	71 	}$	&	127 	$_{	-59 	}	^{+	27 	}$	\\
\noalign{\smallskip}
1999+2011	&	1332-2868	&	133.08 	$_{	-0.007 	}	^{+	0.010 	}$	&	-46.34 	$_{	-0.004 	}	^{+	0.005 	}$	&	3242 	$_{	-34 	}	^{+	35 	}$	&	-$^{\ddagger}$							\\
\hline
\end{tabular}\\
\footnotesize{\textit{$^{\dagger}$The radius displayed is the inner shell radius derived from the SHERPA fit plus half of the non-deconvolved shell width.}\\
\textit{$^{\ddagger}$The deconvolved shell width is not calculated, because the beam FWHM is unclear for this merged data, but an non-deconvolved width of 310$_{-34 }^{+22}$\,arcsec is found.}}
\end{table*}
The optimum 2-dimensional shell fit to the radio continuum suggests a centre of ($\alpha$,$\delta$ J2000)=(133.08$^{\circ}$ $_{-0.007}^{+0.010}$, $-$46.34$^{\circ}$ $_{-0.004}^{+0.005}$) (8h\,52m\,19.2\,s, $-$46$^{\circ}$20$^{\prime}$24.0$^{\prime\prime}$), and a radius of 3242$_{-34}^{+35}$\,arcsec ($\sim$0.9$^{\circ}$). We do not have sufficient coverage in the south-east of the SNR to warrant attempting an egg-shaped functional fit as suggested by X-ray emission studies \citep[e.g. ][]{Aschenbach:1998,Fukui:2017}.

The $\sim$10\,arcsecond-scale errors associated with the SNR radius and the differing coverage of the 1999 and 2011 observations (see Figure\,\ref{fig:ATCAcoverage}) make the data unsuitable for expansion rate studies, particularly because previous X-ray expansion rate measurements \citep{Katsuda:2009,Allen:2015} suggest that sub-arcsecond\,yr$^{-1}$ precision is required for this purpose.

\subsection{Filamentary Features}\label{sec:filaments}
We have identified filamentary structures within the 1999 and 2011 ATCA data-sets that are not aligned with the circular \RX\ shell. We highlight these with dashed lines in Figure\,\ref{fig:Rad_combo}. These features are also indicated on Figure\,\ref{fig:OIII}, which displays [OIII] emission, and Figure\,\ref{fig:GasUV}, which displays short-wavelength UV emission. We note that none of the identified filaments correspond to the TeV gamma-ray emission or hard X-ray emission from the \RX\ shell. The unassociated PWN PSR\,J0855$-$4644 is also indicated in Figure\,\ref{fig:Rad_combo}. Coincident gamma-ray emission at this location is associated \citep{Acero:2013} with PWN PSR\,J0855$-$4644, not \RX .

Filament\,A is a vertical filament of emission at right-ascension $\sim$134$^{\circ}$ 
that partially overlaps with the north-eastern edge of \RX . This feature has previously been identified as `feature\,C' by \citet{Combi:1999}. 
Filament\,B is a $\sim$17$^{\prime}$-length radio continuum structure centred at approximately [133.73,-46.25].  
Filament\,C is a prominent feature of the 1999 and 2011 radio continuum maps which extends almost vertically along right ascension $\sim$132.5$^{\circ}$ for a length of $\sim$1.5$^{\circ}$ inside the perimeter of the \RX\ shell. Filament D is a half-arcminute long filament feature, centred on approximately [133.45,-46.62], at an angle of $\sim$24$^{\circ}$ to the ecliptic plane.

\subsubsection{High Energy Correspondence}\label{sec:highE}
The broad-band X-ray structure in the top right image of Figure\,\ref{fig:Rad_combo} is dominated by thermal emission from the Vela SNR \citep{Aschenbach:1995,Aschenbach:1998}. This means that the broadband X-ray structure can help to distinguish the radio continuum structure of the Vela\,SNR versus Vela\,Jr, allowing us to investigate the origin of Filaments A, B, C and D. Similarly, UV emission in Figure\,\ref{fig:GasUV} highlights the older, foreground Vela SNR with perhaps a smaller contamination from the Vela\,Jr shocks due to the larger distance of Vela\,Jr.

Referring to Figures \ref{fig:Rad_combo} and \ref{fig:GasUV}, filaments A, C and D have corresponding soft X-ray and UV emission. Since this emission is believed to be dominated by thermal emission from the Vela SNR, Filaments A, C and D have clear thermal counterparts, strongly suggesting an association with the Vela SNR. Furthermore, UV and soft X-ray emission towards Filament A appear to extend northwards beyond the \RX\ perimeter, towards more vertical features that are likely associated with the Vela SNR (see Figures \ref{fig:Rad_combo}, \ref{fig:OIII} and \ref{fig:GasUV}). Conversely, all filaments, A, B, C and D, do not have clear correspondences with hard X-rays or gamma-ray emission, suggesting no association with \RX .  

Filament\,B does not have any UV, X-ray or gamma-ray counterpart, so we are unable to favour either a Vela\,SNR or Vela\,Jr origin for this feature.

\subsubsection{Optical Correspondence}
[OIII] emission is considered a good tracer of cooling post-shock gas associated with radiative-phase SNRs, consistent with emission from the Vela\,SNR. In Figure\,\ref{fig:OIII}, the structure of the Vela\,SNR can be seen to overlap and align well with filaments A and D. These have corresponding filamentary [OIII] emission structure, as seen in Figure\,\ref{fig:GasOIII}, which shows [OIII] with [SII] and H$\alpha$ emission. Both filaments B and C have no optical counterpart.

The eastern side of the Filament\,D radio continuum emission is coincident with an arc in [OIII] emission (Figure\,\ref{fig:GasOIII}). This feature, which we have coined the `Vela\,Claw' may be related to a shock interaction with gas, and is discussed in Section\,\ref{sec:gas}. 

Generally, a lack of [OIII] emission towards the outer circular shell of \RX\ suggests that [OIII] does not generally trace the fast $\sim$10$^3$\,kms$^{-1}$ \RX\ shocks, consistent with expectations.

\subsection{Correspondence with the Interstellar Medium Associated with \RX}\label{sec:gas}
\begin{figure}[h]
	\centering
	\includegraphics[width=0.75\columnwidth]{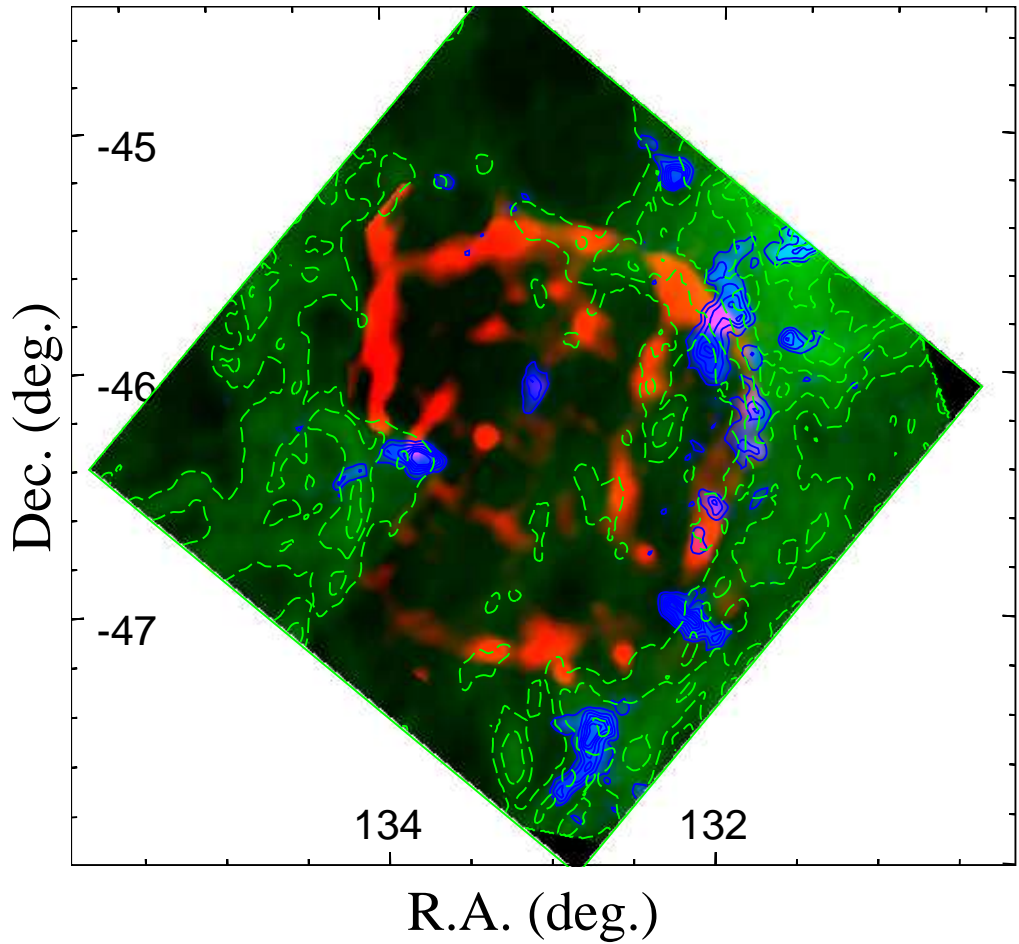}
	\caption{A 3-colour image featuring radio continuum (red, also in Figure\,\ref{fig:Rad_combo}), and gas tracers at the velocity of a void. ATCA and Parkes HI emission integrated between 22 and 33\,km\,s$^{-1}$ is green, with dashed green contour levels of 150, 200 and 250\,K\,km\,s$^{-1}$ is displayed \citep[as seen as Figure\,1 of][]{Fukui:2017}. Nanten $^{12}$CO(1-0) emission integrated between 28 and 33\,km\,s$^{-1}$ is blue, with blue 2, 4, 6, 8, 10 and 12\,K\,km\,s$^{-1}$ contours overlaid \citep[as seen in Figure\,1 of][]{Fukui:2017}. \label{fig:GasRadio}}
\end{figure}

Figure\,\ref{fig:GasRadio} shows the ATCA radio continuum morphology against a backdrop of a HI-dip identified by \citet{Fukui:2017} to be likely associated with \RX . The SNR shell corresponds well to this dip in HI emission (velocity $\sim$30\,kms$^{-1}$), with the eastern side partially-overlapping the contours of atomic gas, while regions of the north-east and south sit outside the lowest HI contour level within the HI-dip. It follows that ATCA radio continuum emission supports the gas-association found by \citet{Fukui:2017}. 

\begin{figure}[h]
	\centering
	\includegraphics[width=0.65\columnwidth]{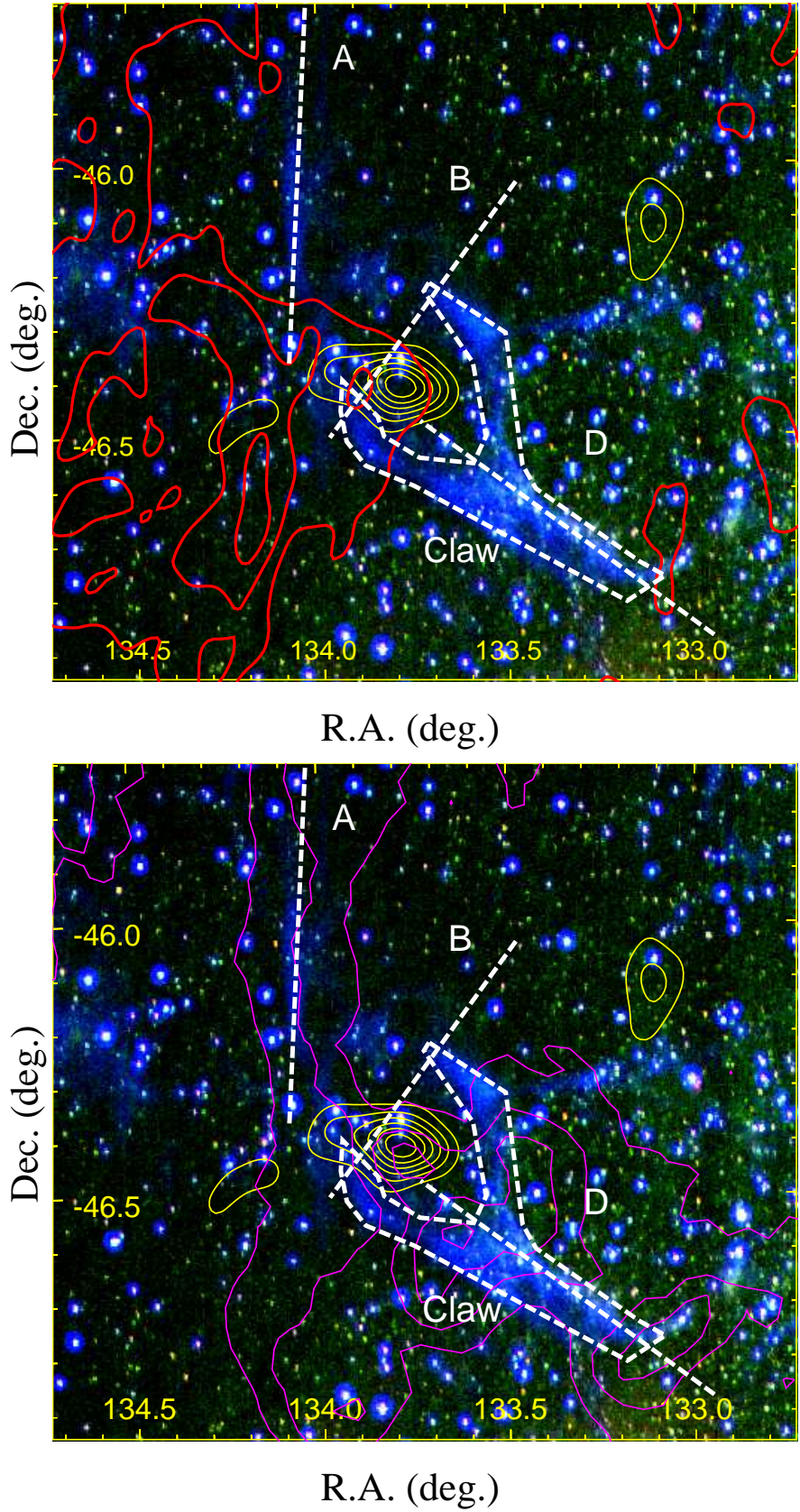}
	\caption{\textbf{Top:} A 3-colour image featuring H$\alpha$ (red), [SII] (green) and [OIII] (blue) from the Boller \& Chivens Telescope. Red 150 and 200\,K\,km\,s$^{-1}$ contours of ATCA and Parkes HI emission integrated between 22 and 33\,km\,s$^{-1}$ are overlaid \citep[as seen as Figure\,1 of]{Fukui:2017}. Yellow 2, 4, 6, 8, 10 and 12\,K\,km\,s$^{-1}$ contours of Nanten $^{12}$CO(1-0) emission integrated between 28 and 33\,km\,s$^{-1}$ are also overlaid \citep[as seen in Figure\,1 of][]{Fukui:2017}. \textbf{Bottom:} Same as top, only red HI contours have been replaced by magenta contours that indicate \textit{ROSAT} broadband (0.2-2.4\,keV) X-ray emission \citep{Aschenbach:1998}. \label{fig:GasOIII}}
\end{figure}

On examination of the [OIII]-traced, Filament\,D, a forking structure was identified: we denote this feature as the Vela Claw and indicate its position and morphology in Figure\,\ref{fig:GasOIII}. The Claw appears to be part of a larger structure that is connected to the [OIII] emission coincident with Filament\, D. The [OIII] filament diverges into two filamentary structures to form a claw-like structure.  
This appears to occur at a location near a stripped CO clump referred to as `CO30E' by \citet{Fukui:2017}, which the authors suggested to be associated with Vela Jr. The diverging filament forms a crescent around clump CO30E from the north-west to the east, suggesting that the [OIII] emission may be associated with the Vela Jr shock. 

Two scenarios are considered to explain the origin of the Vela Claw: (i) Vela\,Jr generating the Vela\,Claw, and (ii) the Vela\,SNR generating the Vela\,Claw. 

Scenario (ii) is consistent with the accepted picture of how optical [OIII] emission is produced from SNRs, i.e. the radiative cooling of diffuse gas in the wake of a $\sim$100\,km$^{-1}$ shockwave (see  Section\,\ref{sec:introVela}). Furthermore the Vela\,SNR is a known emitter of this transition, as is clearly seen extensively throughout the Vela\,SNR shell (see Figure\,\ref{fig:OIII}). It follows that if scenario (ii) is correct, the Vela Claw correspondence with clump CO30E might simply be coincidental. Indeed, this scenario is consistent with the Vela\,Claw being associated with the coincident radio continuum feature, Filament\,D, for which a Vela\,SNR origin is perhaps favoured by the coincident UV/soft X-ray structure attributable to  the Vela\,SNR (see Section\,\ref{sec:highE}). Alternatively, the correspondence might be evidence for an association of clump CO30E with the Vela SNR - a scenario which is disfavoured by the remarkable correlation between gas near the CO30E Galactic velocity and the Vela\,Jr gamma-ray emission \citep{Fukui:2017}, assuming that Vela\,Jr and the Vela\,SNR are indeed at different distances ($\sim$750 and $\sim$300\,pc, respectively), as currently believed.

Scenario (i), an association of the Vela Claw with Vela\,Jr, would be a surprising result because the conditions within the $\sim$3000\,kms$^{-1}$ Vela\,Jr shock are not considered conducive to stimulate optical [OIII] emission, or indeed any significant detectable thermal cooling lines at other wavelengths. Naively, an association of Vela\,Jr with clump CO30E and the Vela\,Claw might require that the SNR shock (forward or reverse) is slowed significantly by the clump CO30E density gradient. If we assume a constant ram pressure model, $P=\rho v^2$, where $\rho$ is the gas density and $v$ is the shock speed, the Vela\,Jr shock might be slowed to $\sim$10\% of the initial speed in a localised region with a sharp density gradient of $\sim$100$\times$. This is plausibly occurring for Vela\,Jr. Clump CO30E has a mass of 180\,M$_{\odot}$ and an approximate radius of 3\,pc \citep[Table 1][]{Fukui:2017}. Assuming a spherical geometry, the average H$_{2}$ density of clump CO30E is approximately n$\sim$30\,cm$^{-3}$, which would represent a $>$100$\times$ increase in density with respect to the density expected for a wind-blown cavity region (e.g. $\sim$0.1\,cm$^{-3}$), like that proposed for the evolution of Vela\,Jr \citep{Fukui:2017}. 

\citet{Sutherland:1995} proposed another mechanism to explain thermal emission from seven young SNRs (including Cassiopeia\,A, SNR G292.0$+$1.8 and Puppis\,A) that may also be able to describe scenario (i) in Vela\,Jr. In the \citeauthor{Sutherland:1995} model, knots of oxygen-rich ejecta material move through a low-density medium until encountering a density discontinuity at a relative velocity of several$\times$1000\,km$^{-1}$. A density increase of 100$\times$ is said to translate to internal cloud shock speeds of $\sim$100\,kms$^{-1}$, which could lead to the observed [OIII] emission.

The validity of scenario (i) is complicated by the expectation that a thermal UV/X-ray emission counterpart to the [OIII] emission might be expected, as is the case for shell segments of the SNR RCW\,86, which exhibits shock velocities that vary by an order of magnitude due to sharp localised density gradients \citep[e.g.][]{Vink:2006_RCW86,Broersen:2014_RCW86}. As discussed in Section\,\ref{sec:highE}, the Vela\,SNR likely dominates the thermal UV/X-ray emission in this region, therefore scenario (i) is difficult to test with UV or X-ray data. Figure\,\ref{fig:GasOIII} (bottom) shows the significant overlap between the Vela\,Claw and \textit{ROSAT} soft thermal X-ray emission. Small-scale spectral studies of the X-ray emission might help to distinguish a thermal emission component from Vela\,Jr towards the Vela\,Claw. Such a detection would not only strengthen the association between the SNR and clump CO30E, but would be further model-dependent evidence of the core-collapse nature of the progenitor event, since the \citeauthor{Sutherland:1995} model requires oxygen-rich ejecta material inherent to core-collapse events.

Also displayed in Figure\,\ref{fig:GasRadio} are CO-traced molecular gas clumps at the outskirts of the HI dip. Towards the north-west region of \RX\ the signal/noise level and resolution of the 2011 ATCA data-set was sufficient across the bands to derive a reliable spectral index map. In Figure\,\ref{fig:SpecIndex}, radio continuum emission towards the rim of \RX\ may flatten ($\gtrsim -$0.5) towards a dense molecular region traced by CO(1-0) emission - a scenario indicative of a SNR-cloud shock interaction \citep[e.g.][]{Keohane:1997,Ingallinera:2014}. No other suggestion of a shock-cloud interaction in this region is seen at other wavelengths, including thermal UV and X-ray emission.

\section{CONCLUSION}
We suggest that the 1332-2868\,MHz radio continuum emission of \RX\ is well-characterised by a 2-dimensional shell of 3242$\pm{35}$\,arcsec centred at (l,b)=(133.08$^{\circ}$ $\pm{0.01}^{\circ}$,-46.34$^{\circ}\pm{0.005}^{\circ}$). Several filamentary structures are identified and multi-wavelength data is examined to investigate their relation to \RX .

Based \textbf{only} on morphological studies of radio continuum, UV, X-ray and gamma-ray emission, three radio filaments towards \RX\ show no indication of an \RX\ origin, while one filament has no clear multi-wavelength counterpart. An investigation of [OIII] emission, however, lead to the identification of a feature we coined the Vela\,Claw, which is possibly associated with one of the radio filaments. The feature corresponds to a south-east molecular clump previously suggested to be stripped by the progenitor of \RX\ \citep{Fukui:2017}. Although the [OIII] feature is consistent with an origin in the shocks of the coincident Vela\,SNR, motivated by morphological correspondence with \RX , we propose the possibility of \RX\ triggering the [OIII] emission of the Vela\,Claw. Proof of an association would reinforce the gas association found by \citet{Fukui:2017}.

\acknowledgments
\section{acknowledgments}
We thank the anonymous referee for their considered and constructive feedback which improved the quality of our manuscript. The Australia Telescope Compact Array is part of the Australia Telescope National Facility which is funded by the Commonwealth of Australia for operation as a National Facility managed by CSIRO. This paper includes archived data obtained through the Australia Telescope Online Archive (http://atoa.atnf.csiro.au). This research has made use of software provided by the Chandra X-ray Center (CXC) in the application packages CIAO, ChIPS, and Sherpa.




%\begin{thebibliography}{}
\bibliographystyle{aasjournal}
\bibliography{ReferencesVelaJr}

\end{document}